\documentclass[sn-standardnature, iicol]{sn-jnl}
\usepackage{lineno}
\let\oldequation\equation
\let\oldendequation\endequation

\renewenvironment{equation}
  {\linenomathNonumbers\oldequation}
  {\oldendequation\endlinenomath}

\jyear{2023}%

\theoremstyle{thmstyleone}%
\theoremstyle{thmstyletwo}%

\theoremstyle{thmstylethree}%

\begin{document}

\title[Article Title]{Vernier Microcombs for Integrated Optical Atomic Clocks}


\author*[1]{\fnm{Kaiyi} \sur{Wu}}\email{\small wu1871@purdue.edu} \equalcont{\small These authors contributed equally to this work.}

\author[1]{\fnm{Nathan} \sur{P. O'Malley}} \equalcont{\small These authors contributed equally to this work.}

\author[1]{\fnm{Saleha} \sur{Fatema}}
\author[1,3]{\fnm{Cong} \sur{Wang}}
\author[2]{\fnm{Marcello} \sur{Girardi}}
\author[4]{\fnm{Mohammed} \sur{S. Alshaykh}}
\author[2]{\fnm{Zhichao} \sur{Ye}}
\author[1,5]{\fnm{Daniel} \sur{E. Leaird}}
\author[1]{\fnm{Minghao} \sur{Qi}}
\author[2]{\fnm{Victor} \sur{Torres-Company}}
\author[1]{\fnm{Andrew} \sur{M. Weiner}}

\affil[1]{\small\orgdiv{School of Electrical and Computer Engineering}, \orgname{Purdue University}, \orgaddress{\city{West Lafayette}, \postcode{47907}, \state{IN}, \country{USA}}}

\affil[2]{\small\orgdiv{Department of Microtechnology and Nanoscience}, \orgname{Chalmers University of Technology}, \orgaddress{\street{SE-41296},  \country{Sweden}}}

\affil[3]{\small\orgdiv{Currently with Department of Surgery}, \orgname{University of Pittsburgh}, \orgaddress{\postcode{15219}, \state{PA}, \country{USA}}}

\affil[4]{\small\orgdiv{Electrical Engineering Department}, \orgname{King Saud University}, \orgaddress{\city{Riyadh}, \postcode{11421}, \country{Saudi Arabia}}}

\affil[5]{\small\orgdiv{Currently with Torch Technologies}, \orgname{supporting AFRL/RW}, \orgaddress{\city{Eglin Air Force Base}, \state{FL}, \country{USA}}}


\abstract{

CMOS-compatible Kerr microcombs have drawn substantial interest as mass-manufacturable, compact alternatives to bulk frequency combs. This could enable deployment of many comb-reliant applications previously confined to laboratories. Particularly enticing is the prospect of microcombs performing optical frequency division in compact optical atomic clocks. Unfortunately, it is difficult to meet the self-referencing requirement of microcombs in these systems due to the $\sim$THz repetition rates typically required for octave-spanning comb generation. Additionally, it is challenging to spectrally engineer a microcomb system to align a comb mode with an atomic clock transition with sufficient signal-to-noise ratio. Here, we adopt a Vernier dual-microcomb scheme for optical frequency division of a stabilized ultranarrow-linewidth continuous-wave laser at 871 nm to a $\sim$235 MHz output frequency. In addition to enabling measurement of the comb repetition rates, this scheme brings the freedom to pick comb lines from either or both of the combs. We exploit this flexibility to shift an ultra-high-frequency ($\sim$100 GHz) carrier-envelope offset beat down to frequencies where detection is possible and to place a comb line close to the 871 nm laser - tuned so that if frequency-doubled it would fall close to the clock transition in $^{171}$Yb$^+$. Moreover, we introduce a novel scheme which suppresses frequency noise arising from interferometric phase fluctuations in our dual-comb system and reduces the frequency instability down to our measurement limit. Our dual-comb system can potentially combine with an integrated ion trap toward future chip-scale optical atomic clocks. 
}




\maketitle


\begin{figure*}[h]%
\centering
\includegraphics[width=0.9\textwidth]{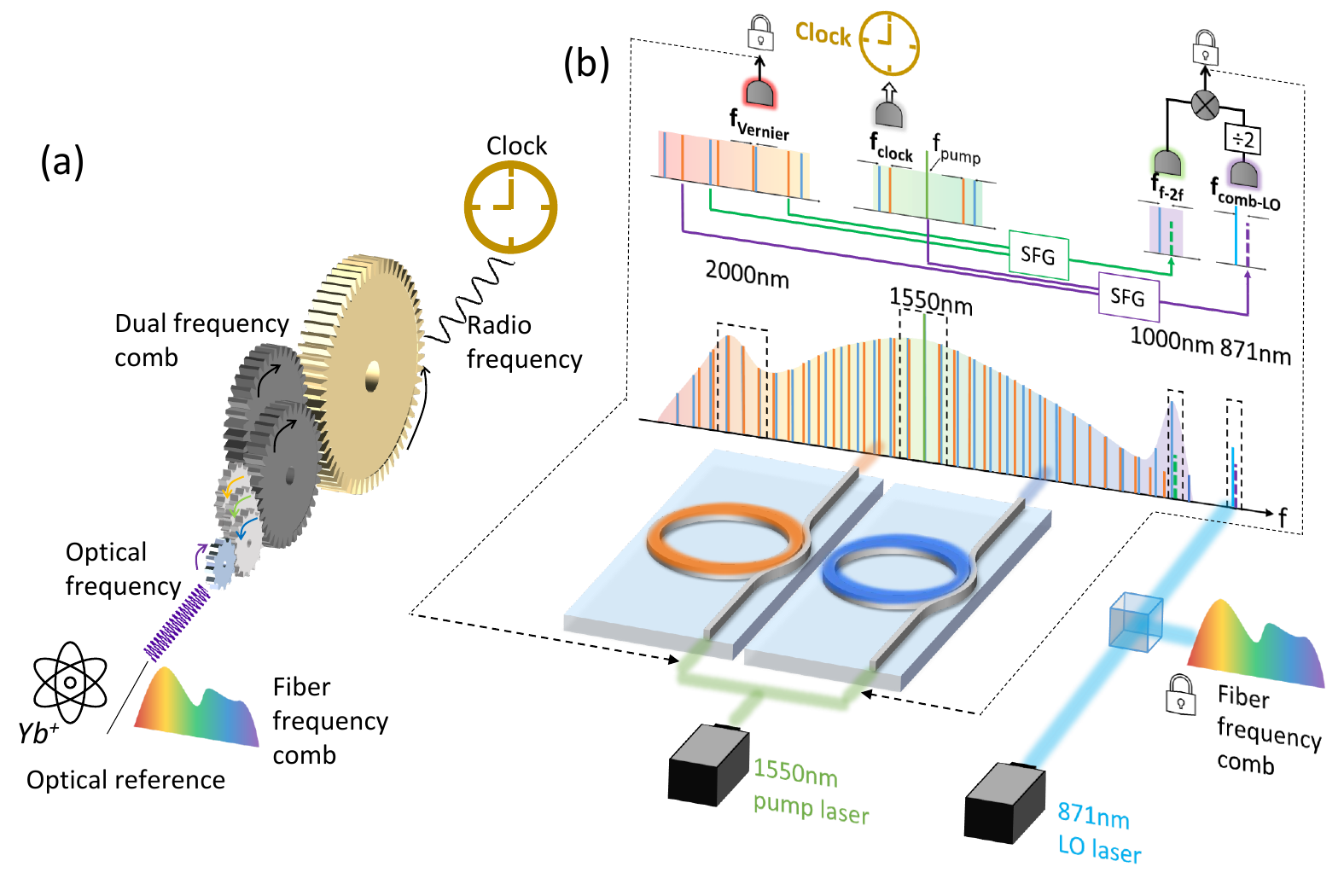}
\caption{\textbf{a,} An optical frequency comb system is analogous to a gear set, transferring the stability from optical to radio frequency. The large (small) size gears represent radio (optical) frequencies. The optical reference can be a Yb$^+$ ion trap (as our system is designed for) or a stabilized fiber comb (as a stable frequency proxy we adopted). \textbf{b,} Illustration of a Vernier dual-microcomb system for optical frequency division. The dual combs excited by a shared pump at 1550 nm generate broadband spectra spanning from $\sim$1 $\mu$m to $\sim$ 2 $\mu$m. The dashed boxes indicate the four spectral regions being employed, and the zoom-in views are shown above. The first sidebands around the pump are used for RF clock output. The $\sim$ 2 $\mu$m light is used for Vernier beat detection and SFG process (green arrow line) for f-2f. The dual-comb is related to the 871 nm LO laser through summing the pump and one of the Vernier comb lines at $\sim$ 2 $\mu$m (purple arrow line), and the 871 nm laser is locked to a stabilized frequency comb. The black dotted lines indicate the two locks feedback to the pump intensities of the two combs. }\label{fig1}
\end{figure*}

By virtue of their extreme long-term frequency stability, optical atomic clocks promise to revolutionize timing systems, enable fundamental tests of physics \cite{godun2014frequency}, and allow chronometric leveling and geodesy \cite{takano2016geopotential, riehle2017optical}. Lab-scale optical lattice clocks \cite{campbell2017fermi, collaboration2021frequency} and ion trap clocks \cite{brewer2019al+, collaboration2021frequency} can provide state-of-the-art frequency stability.
However, scaling these clocks to a low size, weight, and power (SWaP) architecture is an important and challenging hurdle preventing significant deployment of the highest-performing optical atomic clock technology. Efforts such as \cite{ivory2021integrated, ropp2023integrating} toward integrated ion traps and optical lattices illustrate recent progress toward high-stability, chip-scale optical atomic clocks.

Microcombs are an essential part of future chip-scale optical atomic clocks \cite{papp2014microresonator, newman2019architecture} as they establish a coherent link from atomic references at $\sim$ hundreds of THz all the way down to the RF domain (analogous to a gear set - Fig. \ref{fig1}a \cite{diddams2020optical}) while preserving the stability of an atom-referenced clock laser. The realization of this coherent link is referred to as optical frequency division (OFD).
The comb's optical modes $f_{\rm m}$ are equally spaced by the repetition rate $f_{\rm rep}$ (usually an RF frequency), and the whole comb has a spectral offset (the carrier-envelope offset [CEO] frequency, $f_{\rm CEO}$) so that $f_{\rm m}=mf_{\rm rep}+f_{\rm CEO}$. The RF mode spacing of optical modes in a comb is key to enabling the bridge between optical and RF domains. Historically frequency combs in optical atomic clocks have often been generated by modelocked lasers \cite{diddams2001optical, ye2001molecular} that have volumes on the order of liters, and generally require at least a measure of hand assembly, complicating mass production and increasing costs.
Shrinking down these combs should be helpful toward miniaturizing optical atomic clocks and other systems requiring stabilized combs, and can be achieved through frequency comb generation in Kerr microresonators on complementary metal-oxide semiconductor (CMOS)-compatible photonic chips \cite{kippenberg2018dissipative}.

Unfortunately, Kerr microcombs possessing the octave span needed for self-referencing (a requirement for many OFD applications) typically have $\sim$THz repetition rates. This leads to $f_{\rm CEO}$ in the range of $\pm 500$ GHz, which is difficult to control because of microring fabrication imperfections \cite{yu2019tuning, moille2021tailoring}; as a result, the $f_{\rm CEO}$ beat is frequently too high for electronic detection. 
In part due to these issues, microcomb-based OFD remains a challenging task. 
The only two demonstrations \cite{papp2014microresonator, newman2019architecture} of microcomb-based OFD of an atom-referenced laser, to our knowledge, relied on low repetition rate (tens of GHz), narrowband combs generated from mm-size whispering gallery mode silica microdisk resonators, either spectrally broadened in a nonlinear fiber \cite{papp2014microresonator} or used in conjunction with a secondary, broadband silicon nitride THz-microcomb \cite{newman2019architecture}. Both of these works utilized thermal vapors for the atomic reference, not cooled or trapped atoms.

In this work, we demonstrate an integrated photonic platform based on Vernier dual-microcombs that overcomes some of the fundamental challenges of previous microcomb-based systems for the realization of chip-scale optical clocks. Specifically, by pairing a main octave-spanning $\sim$THz-microcomb with a secondary broadband $\sim$THz Vernier microcomb, both on a silicon nitride (SiN) platform, we successfully frequency divide an ultranarrow-linewidth CW laser at 871 nm \cite{lai2021871nm} to an RF clock output at $\sim$235 MHz. 
This laser is designed for frequency-doubling to within a few GHz of the Ytterbium ion ($^{171}$Yb$^+$) ``clock transition" at 435.5 nm that is expected to support better frequency stability than most thermal atomic references. By virtue of its dual broadband combs, the Vernier scheme used here presents a substantial advantage over single broadband comb schemes.
One of the advantages is the freedom of picking comb lines from either or both of the combs for frequency-summing to aid in reaching a much greater variety of wavelengths \cite{Wu2023vernier}.
Here this is adopted both to help circumvent the high-frequency $f_{\rm CEO}$ detection problem and to reach 871 nm by frequency-summing the pump and one of the secondary comb's lines at $\sim$2$\mu$m. While we focus on this transition, we emphasize the generality of the technique to reach a variety of other atomic transitions for future optical clocks.
In addition, with both THz-microcombs fabricated in SiN films of the same thickness, there is potential to integrate our scheme on a single chip. Since the area of a microring scales as the square of its diameter (inverse square of the comb repetition rate), the footprint of a dual THz-microcomb structure could be several orders of magnitude smaller than that of architectures employing GHz microcombs. Furthermore, the all-planar geometry employed here avoids the complexity involved in coupling to suspended whispering gallery mode resonators or integrating the same \cite{yang2018bridging}. These results demonstrate a versatile and general microcomb platform for the realization of chip-scale optical atomic clocks. 

\vspace{\baselineskip}
\noindent{\textbf{Results}}

\begin{figure*}[h]%
\centering
\includegraphics[width=0.9\textwidth]{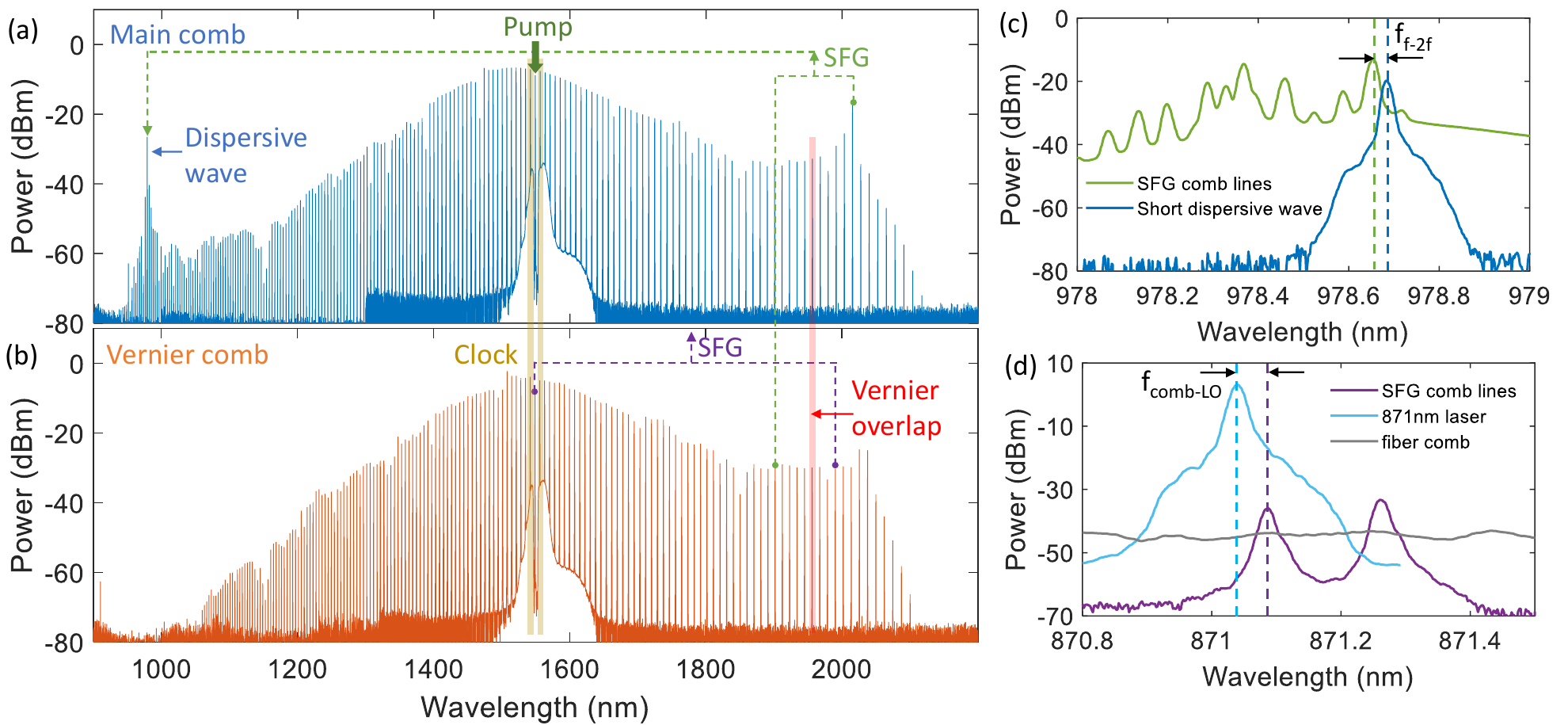}
\caption{\textbf{a-d} Optical spectra of \textbf{a}, the main comb and \textbf{b}, the Vernier comb, each measured after pump suppression using a coarse wavelength-division multiplexing filter, \textbf{c}, the short wavelength main comb line (dispersive wave) (blue) and the SFG lines at 1 $\mu$m (green) for f-2f beat detection, and \textbf{d}, the 871 nm LO laser (light blue), SFG at 871 nm (purple), and the fiber comb lines (grey). The fiber comb lines are spaced by $\sim$250 MHz, which is smaller than the resolution of the optical spectrum analyzer ($\sim$1 GHz). Hence, individual comb lines are not resolved here, and the trace appears featureless. \text{c-d} are measured at 0.01 nm resolution.}\label{fig2}
\end{figure*}

\noindent{}\textbf{Dual-microcomb scheme.} The overview of our experiment configuration is visualized in Fig. \ref{fig1}b. An ultra-narrow linewidth 871 nm ($\sim$344 THz) continuous wave (CW) laser (LO laser) developed by OEwaves \cite{lai2021871nm, lai2022ultra} is phase-locked to an external self-referenced fiber comb (FC, Menlo systems), which here serves as a frequency-stable proxy for a potential Yb$^+$ frequency reference (which is not available in our laboratory). Our microcomb system is built on the Vernier dual-microcomb scheme \cite{wang2020vernier, Wu2023vernier}, which is leveraged here for full optical frequency division for the first time. 
Two broadband microcombs with large repetition rates ($\sim$900 GHz), slightly offset from one another ($\delta f_{\rm rep} \sim$ 20 GHz), are generated with a shared CW pump. This configuration has been shown to allow detection of both large $f_{\rm rep}$ and $f_{\rm CEO}$ \cite{Wu2023vernier}. We aim to transfer the LO laser stability to an RF clock output using a heterodyne beat in the dual-microcomb system.  

We utilize a pair of microrings with $\sim$25 $\mu$m radius, fabricated in SiN \cite{ye2019high}, to generate our microcombs. The main comb is designed to have dispersive waves at $\sim$1 $\mu$m and $\sim$2 $\mu$m wavelengths to assist the f-2f process, while the Vernier comb should have a resonance wavelength close to that of the main comb to allow for a shared pump. We select microrings fabricated on separate chips but the same wafer. A single CW laser around 1550 nm is amplified and split to pump both the main and Vernier devices simultaneously, generating an octave-spanning coherent frequency comb in each. Figure \ref{fig2}a,b show the spectra for the main and Vernier combs. The main (Vernier) comb possesses a repetition rate $f_{\rm rep1} \sim$896 GHz ($f_{\rm rep2} \sim$876 GHz). As a result of the shared pump, when the first sidebands from each comb are heterodyned together on a photodetector, they create an electronic signal oscillating at the difference of the two repetition rates (see Fig.\ref{fig1}b). We call this signal $f_{\rm clock}$, where
\begin{equation}
    f_{\rm clock} = f_{\rm rep1} - f_{\rm rep2} \approx 19.7 GHz,
    \label{eq clock}
\end{equation}
and through appropriate feedback to our two combs, seek to transfer the stability of the 344 THz LO laser to this $\sim$19.7 GHz dual-comb beat. 

In order to accomplish OFD from the LO laser to $f_{\rm clock}$, at a minimum, we need to stabilize the two repetition rates to the LO laser, and thus several optical signals must be photodetected and stabilized. We separate the 1 $\mu$m, 1.55 $\mu$m and 2 $\mu$m spectral components and combine the dual-comb spectra at 1.55 $\mu$m and 2 $\mu$m wavelengths for the detection of various heterodyne beats. The spectral domain schematic is illustrated in Fig. \ref{fig1}b. A detailed experimental setup can be found in Methods. Due to their different repetition rates, the two combs walk off from each other as one moves away from the pump, eventually coming back together at the so-called Vernier overlap point. The comb modes at this point can be photodetected (here, around 1956 nm) to produce a relatively low-frequency beat:
\begin{equation}
    f_{\rm Vernier} = 44 f_{\rm rep1} - 45 f_{\rm rep2} \approx -8 GHz. 
    \label{eq Vernier}
\end{equation}
With this signal phase-locked through feedback to the Vernier comb, $f_{\rm rep2}$ will follow $f_{\rm rep1}$ \cite{omalley2022vernier}. We still need to stabilize the two repetition rates to the LO laser through feedback to the main comb to achieve a fully stabilized $f_{\rm clock}$, the difference of the two repetition rates.

We perform a novel f-2f process via sum-frequency generation (SFG) \cite{Wu2023vernier} in a periodically-poled lithium niobate (PPLN) waveguide, which sums one comb line from each comb at around 2 $\mu$m ($f_{\rm CEO1}+166f_{\rm rep1} \& f_{\rm CEO2}+179f_{\rm rep2}$). The sum products are combined with the short-wavelength side of the main comb around 1 $\mu$m ($f_{\rm CEO1}+342f_{\rm rep1}$) to create an f-2f-like beat signal, $f_{\rm f-2f}$. Fig. \ref{fig2}c shows the sample spectra of the sum products and the short wavelength line.  
This f-2f beat contains contributions from not only the CEO frequency of one of the combs but also the two repetition rates: 
\begin{equation}
    f_{\rm f-2f} = f_{\rm CEO2} - 176f_{\rm rep1} + 179f_{\rm rep2} \approx 11 GHz.
    \label{eq f-2f}
\end{equation}

Finally, to relate the dual-microcombs to the stable 871nm LO laser, which is outside the span of both comb spectra, we use a second PPLN waveguide for SFG between the 1550nm-pump ($f_{\rm CEO2}+220f_{\rm rep2}$) and a long-wavelength comb line ($f_{\rm CEO2}+171f_{\rm rep2}\sim$1990 nm) from the Vernier comb to create a nonlinear product. The LO laser at $f_{871}$ is tuned to $\sim$871.042nm to beat against the SFG product to create $f_{\rm comb-LO}$, and we have 
\begin{equation}
    f_{\rm comb-LO} = 2f_{\rm CEO2} + 391f_{\rm rep2} - f_{871} \approx -19 GHz. 
    \label{eq comb-LO}
\end{equation}
The sample spectra for the SFG lines and the LO laser can be seen in Fig. \ref{fig2}d. This SFG process benefits from the high power of the pump and the optically amplified 2 $\mu$m comb line, leading to a reasonable SFG power of $\sim$-36 dBm. We note that as a result of the common pump condition ($f_{\rm pump}=f_{\rm CEO1}+216f_{\rm rep1}=f_{\rm CEO2}+220f_{\rm rep2}$), the $f_{\rm f-2f}$ and $f_{\rm comb-LO}$ beats in Eq. \ref{eq f-2f}, \ref{eq comb-LO} can be expressed in terms of either $f_{\rm CEO1}$ or $f_{\rm CEO2}$. 

As both $f_{\rm f-2f}$ and $f_{\rm comb-LO}$ contain the $f_{\rm CEO2}$ term, but with a fixed ratio of 1:2, we can cancel the CEO term through electronic frequency division and frequency mixing:
  \begin{equation}
  \begin{split}
      f_{\rm xCEO} &= f_{\rm f-2f} - \frac{f_{\rm comb-LO}}{2} \\
      &= \frac{f_{\rm 871}}{2} - 176f_{\rm rep1} - 16.5f_{\rm rep2}.
      \label{eq xCEO}
    \end{split}
  \end{equation}
We refer to the frequency of the resulting signal as $f_{\rm xCEO}$, since ideally, contributions from both combs' $f_{\rm CEO}$ are removed. Instead, $f_{\rm xCEO}$ depends on the two combs' repetition rates and relates them to the LO laser. If we lock this signal through feedback to the main comb and simultaneously lock  $f_{\rm Vernier}$ by feedback to the Vernier comb, the repetition rates will each be stabilized to the LO laser.
Furthermore, as a consequence of the stabilized repetition rates and the common pump frequency, $f_{\rm clock}$ should be stabilized as well, completing the clock frequency division. Importantly, this is irrespective of the free-running CEO frequencies of both combs. This simplifying scheme allows us to construct our clock system with two servos, removing requirements for a third servo to stabilize the offset frequency. In the experiment, we generate $f_{\rm xCEO}/8$ due to electronic component availability. 

\begin{figure*}[h]%
\centering
\includegraphics[width=0.9\textwidth]{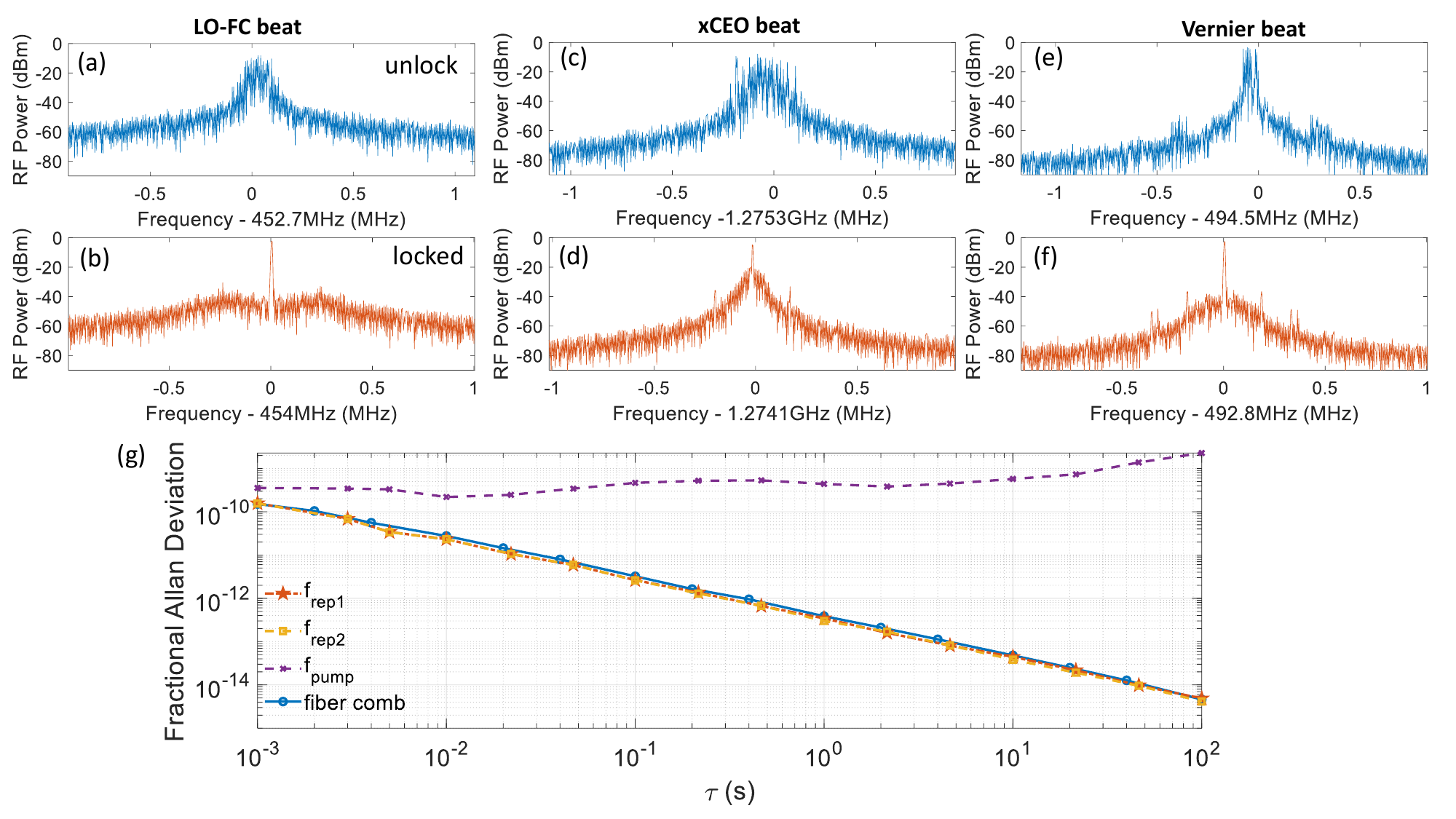}
\caption{\textbf{a-f} ESA traces of the three beats for stabilization \textbf{a-b}, $f_{\rm LO-FC}$, \textbf{c-d}, $f_{\rm xCEO}/8$ and \textbf{e-f}, $f_{\rm Vernier}/8$ when they are unlocked (blue traces) and phase-locked (orange traces). These traces are measured at the lockbox monitor outputs, and the beats are divided by an additional factor of 2. The resolution bandwidth is 3 kHz and the spectral span is 2 MHz. \textbf{g}, Fractional Allan deviation of the repetition rates of the main (orange) and the Vernier (yellow) combs, the pump laser (purple), and the fiber comb reference's repetition rate [and hence an estimate of its fractional optical stability] (blue). }\label{fig3}
\end{figure*}

\vspace{\baselineskip}
\noindent{}\textbf{Microcomb Stabilization.} To achieve the clock frequency division from the optical reference, we first stabilize the 871 nm LO laser to one optical line from a stable fiber comb $f_{\rm FC}$ through an offset phase lock referenced to an RF oscillator, where the offset frequency $f_{\rm LO-FC}\sim \pm$910 MHz is defined as
\begin{equation}
    f_{\rm LO-FC}= f_{871} - f_{\rm FC}.
    \label{eq LO-Menlo}
\end{equation} 
We then phase lock the $f_{\rm xCEO}$ and $f_{\rm Vernier}$ beats to two other RF oscillators by feedback to the pump power of the main and Vernier combs, respectively, using intensity modulators before input-coupling to the two microrings. This stabilization scheme without CEO frequency locking avoids tuning the pump frequency and intensity simultaneously, which may introduce crosstalk between repetition rate and CEO frequency control \cite{del2016phase}. Figures \ref{fig3} a-f show the electrical spectrum analyzer (ESA) traces of the three beats before and after phase-locking. The three oscillators for the offset locks, the fiber comb, and the frequency stability testing utilities are all synchronized to a common GPS-disciplined high-stability ovenized quartz oscillator (OCXO) to eliminate relative drifting. In principle, this sync signal could instead be derived from an atomic reference, as in \cite{diddams2001optical}.

To verify the repetition rates for both microcombs are stabilized to the optical reference (fiber comb), we conduct out-of-loop measurements using an electro-optic (EO) frequency comb \cite{metcalf2013high} to downshift the repetition rates to an electronically detectable range. The time traces of the repetition rates are recorded by a zero dead-time frequency counter running at 1 ms gate time. We calculate the fractional Allan deviation of the two repetition rates, as shown in Fig. \ref{fig3}g. The frequency stability of the fiber comb is also obtained by measuring its $\sim$250 MHz-repetition-rate beat note using a phase noise test set (PNTS), as the blue trace in Fig. \ref{fig3}g shows. This is an estimate of the fractional frequency instability of the optical modes of the fiber comb (see supplementary).

Both microcomb repetition rates follow the optical reference, indicating the stability of the LO laser has been successfully transferred to the two $\sim$THz repetition rates. Note that neither microcomb is fully stabilized because the CEO frequencies of both combs are still free-running. We demonstrate this by beating the pump line with a fiber comb line at $\sim$1550 nm and recording the output with a frequency counter.  The result is shown as the purple trace in Fig. \ref{fig3}g, which has frequency instability that increases with increased averaging time ($\tau$).  Clearly, without stabilizing the CEO frequencies, individual microcomb lines drift over time. Nevertheless, with the stabilized repetition rates, we anticipate a frequency division effect from the optical reference to $f_{\rm clock}$, following the expression derived from the relations between the detected beats (see Supplementary for details) 
\begin{equation}
\begin{split}
   f_{\rm clock} &= \frac{1}{17292}(f_{\rm 871}-2f_{\rm xCEO}+385f_{\rm Vernier})\\
   & \approx \frac{1}{17292}f_{\rm 871}.\\
   \label{eq freqdiv}
   \end{split}
\end{equation}
The fractional frequency instability of the clock output also matches that of the optical reference as a result. 

\vspace{\baselineskip}
\noindent \textbf{RF Clock Performance.} We characterized the fractional frequency instability of the RF clock output with the PNTS. The results are shown in Fig. 4a. Although initially, the fractional instability (orange line, “initial clock”) does come down approximately inversely with $\tau$ (as expected), it shows excess noise with $\sim$5$\times$ higher fractional instability than our system floor (blue line, “fiber comb”) and plateaus at a value of $\sim 10^{-12}$ for $\tau \approx$ 1 sec and beyond. A similar effect has been reported in other recent works \cite{zang2020millimeter, drake2019terahertz} but was not  investigated therein.  We attribute this excess noise to time-varying phase perturbations in the fiber leads that bring light in and out of the individual microrings. Related noise effects are well known in the distribution of ultra-stable frequency references and coherent optical carriers over fiber links and are usually addressed by feedback control to an acousto-optic (AO) frequency shifter in order to stabilize a heterodyne beat involving two-way propagation through both the AO and the fiber link \cite{foreman2007remote, newbury2007coherent}. Here we introduce a novel open-loop electronic mixing technique that reduces the excess noise from our clock signal without the need for two-way propagation and brings the frequency instability down to a level near our system floor, explained below.

\begin{figure*}[h]%
\centering
\includegraphics[width=0.9\textwidth]{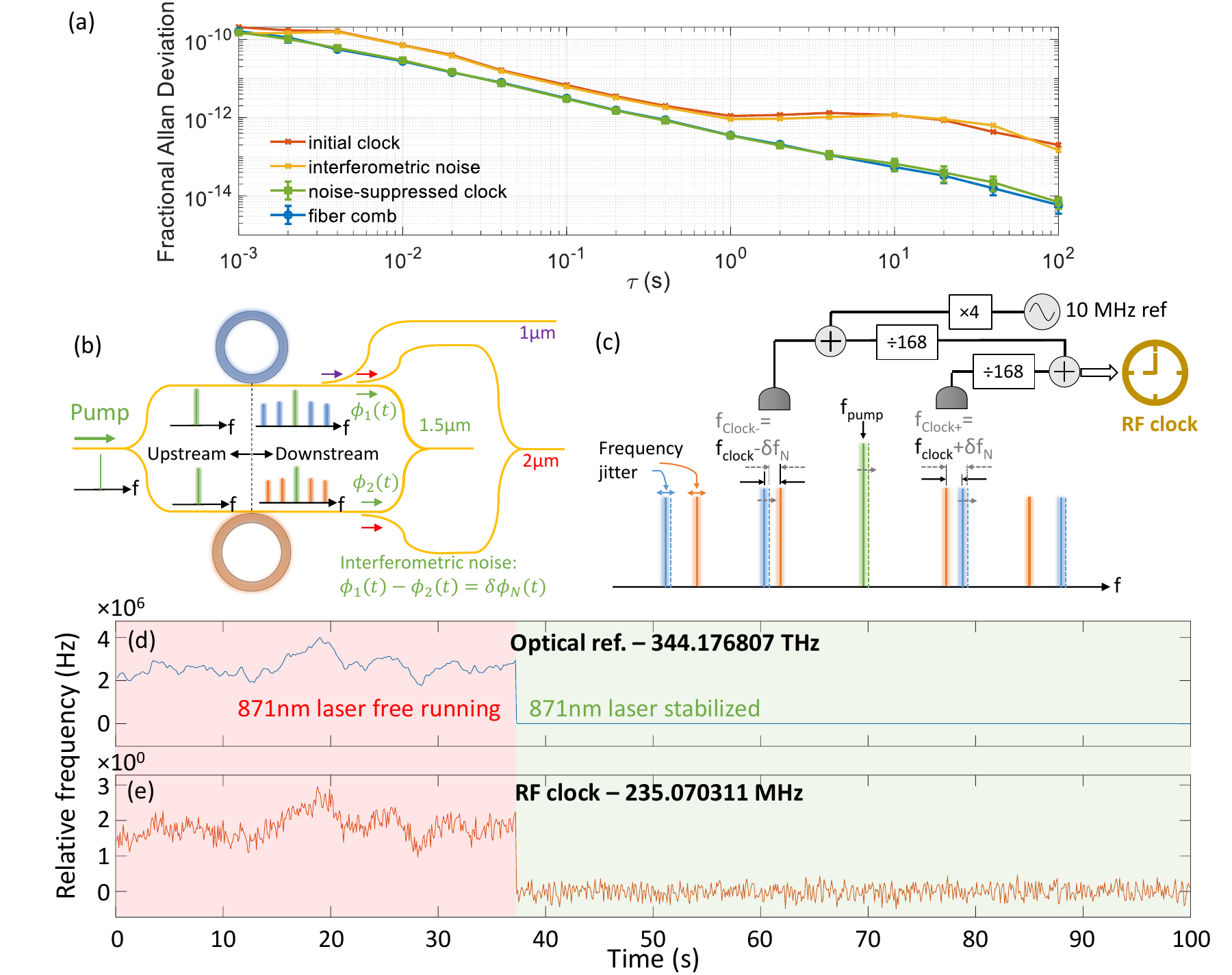}
\caption{\textbf{a}, fractional Allan deviation of the clock without (orange) and with (green) noise suppression, the extracted interferometric noise (yellow) and the optical reference (blue). \textbf{b}, Simplified dual-comb setup portraying the interferometric noise of this configuration.  The blurring of the different comb lines represents frequency jitter due to interferometric phase noise. \textbf{c}, Noise suppression scheme exploiting dual clock signals generated using distinct photodetectors driven by main-Vernier comb line pairs at frequencies higher and lower than the pump frequency. The dashed comb lines illustrate a hypothetical case when the pump to the main ring (green line) and the resulting main ring comb lines (blue) are shifted up in frequency by time-varying phase $\phi_1 (t)$, while the pump to Vernier ring and resulting Vernier comb lines are unaffected [$\phi_2 (t)$ = 0]. The frequency noise of the RF beats generated from the higher and lower comb line pairs is strongly correlated but of opposite signs.  Therefore, by summing the two clock signals with an electronic mixer, the interferometric frequency noise is largely suppressed. \textbf{d-e}, frequency counter traces of \textbf{d} the 871 nm laser and \textbf{e} the RF clock with noise suppression. The 871nm laser is initially free-running, then is locked at around 40 seconds. The $f_{\rm xCEO}/8$ and $f_{\rm Vernier}/8$ locks are on throughout the measurement. The y-axis spans of the optical reference and the RF clock differ by their scaling factor of 17292$\times$84. }\label{fig4}
\end{figure*}

A simplified dual-comb configuration is sketched in Fig. \ref{fig4}b, where a single pump is split into two paths for generating both combs, and the comb lines from both combs are combined together to generate a heterodyne beat signal for the RF clock output. This structure resembles an interferometer, which is sensitive to phase perturbations in either of its arms. We represent the phase fluctuations in the upper and lower arms as $\phi_1 (t)$ and $\phi_2 (t)$, respectively. $\phi_1 (t)$ is comprised of the time-varying phase experienced by the pump in the interferometer arm upstream of the main ring plus the phase picked up by the comb line of interest downstream from the main ring; $\phi_2 (t)$ is defined similarly but refers to the leads connecting the Vernier ring. The heterodyne beat $f_{\rm clock}$ can be generated by photodetecting the first sidebands at either side of the pump, as illustrated in Fig. \ref{fig4}c. We term these $f_{\rm clock+}$ and $f_{\rm clock-}$, corresponding to the higher and lower frequency side of the pump, respectively:
\begin{equation}
\begin{split}
    f_{\rm clock+} &= [f_{\rm pump} + f_{\rm rep1}+\frac{\phi_1'(t)}{2\pi}] \\
    & - [f_{\rm pump} + f_{\rm rep2}+\frac{\phi_2'(t)}{2\pi}], \label{eq clock+}\\
\end{split}
\end{equation}
\begin{equation}
\begin{split}
    f_{\rm clock-} &= [f_{\rm pump} - f_{\rm rep2}+\frac{\phi_2'(t)}{2\pi}] \\
    &- [f_{\rm pump} - f_{\rm rep1}+\frac{\phi_1'(t)}{2\pi}], \label{eq clock-} \\
\end{split}    
\end{equation}
where $\phi'(t)$ represents the time-derivative of $\phi(t)$. These expressions take into account the frequency fluctuations that arise in proportion to the derivative of the time-varying phases \cite{weiner2011ultrafast}. Since $\phi_1 (t)$ and $\phi_2 (t)$ may be uncorrelated, they give rise to frequency noise on the clock signal. Critically, the frequency noise on $f_{\rm clock+}$ and $f_{\rm clock-}$ is equal and opposite in the expressions above. By summing $f_{\rm clock+}$ and $f_{\rm clock-}$ with an electronic mixer
\begin{equation}
    f_{\rm clock+} + f_{\rm clock-} = 2(f_{\rm rep1} - f_{\rm rep2}),
\end{equation}
we can ideally suppress the frequency noise completely. Note that the formulation above ignores any noise contributed by the electronics and approximates the phase shifts incurred by comb lines ±1 in the fiber leads downstream of the resonators as equal. Since the frequencies of the ±1 modes differ from the pump frequency by only a small fractional amount, the error involved in this approximation is small.

We implement this differential noise suppression scheme by detecting $f_{\rm clock+}$ and $f_{\rm clock-}$ on separate photodetectors in parallel and frequency-mixing the resulting heterodyne beats (see Fig. \ref{fig4}c). In practice, we mix a frequency-divided $f_{\rm clock+}/168$ and a copy of $f_{\rm clock-}$ upshifted by 40 MHz and subsequently frequency-divided by 168 (see Fig \ref{fig4}c - more details in Methods). The result is a noise-suppressed RF clock output at $\sim$235 MHz. 

To demonstrate frequency division, we measure the frequency of the optical reference $f_{871}$ and the RF clock output as a function of time using two synchronized counters with the two repetition rate locks in place.  The results are plotted in Figs. 4d,e.  The y-axis span of the optical reference is exactly 17292$\times$84 times that of the RF clock, corresponding to the optical frequency division factor of 17292 (Eq. \ref{eq freqdiv}) and the electronic division factor of 84 (= 168/2) for bringing down the clock frequency to within counter and electronic mixer ranges. The 871 nm LO laser is initially free-running, and we switch on the lock to the fiber comb midway through the measurement.  The RF clock frequency follows that of the LO laser, both when the 871 nm laser is free-running and when it is locked.   Further details, including a discussion of the fluctuations in the RF clock trace, can be found in the Supplement.

The fractional Allan deviation of our $\sim$235 MHz noise-suppressed RF clock output, measured using the PNTS, is plotted as part of Fig. \ref{fig4}a (green trace).  The frequency instability is substantially reduced, now essentially overlaying with the estimated fractional Allan deviation of the frequency comb reference for the LO laser, which determines our measurement floor.  These results signify that the frequency stability of the 344 THz LO laser has been successfully transferred to our $\sim$235 MHz RF clock.  Our method also allows us to extract the differential-mode interferometric frequency noise directly (see Methods), which we also plot in Fig.\ref{fig4}a (yellow trace). It roughly overlays with the red trace representing the fractional frequency instability of the initial clock output.  This supports our attribution of the excess noise present in the clock prior to the suppression scheme to differential mode interferometric phase variations.

\vspace{\baselineskip}
\noindent{\textbf{Discussion}}

\noindent We have demonstrated the use of a Vernier dual-microcomb system to frequency-divide a compact, ultra-narrow linewidth CW laser at 871 nm down to an RF output of $\sim$235 MHz using only two feedback servos, enabled in part by an interferometric noise suppression scheme. The CW laser at 871 nm is within a few GHz of being able to be frequency-doubled to a clock transition of a Ytterbium ion.  

The Vernier dual-microcomb OFD platform used here is useful for detecting high-frequency $f_{\rm CEO}$ and potentially for performing OFD on a variety of atomic species through a dual-comb sum-frequency process. With the continued advancement of microcomb system integration capabilities \cite{rao2021towards}, in the future, our dual combs can potentially be integrated onto a single SiN chip, together with on-chip thermal heaters for spectral alignment and microcomb feedback, and spectral filtering for separation and routing of different wavelength bands. Furthermore, there has been exciting progress in heterogeneous integration of III-V \cite{xiang2021laser, xiang2023three} and thin-film lithium niobate \cite{churaev2023heterogeneously, ruan2023high} materials on the SiN platform. These may enable integration of III-V lasers for microcomb pumping and PPLN for second-order nonlinear frequency conversion towards a fully integrated dual-comb system in the future. Additionally, through the elimination of many of the fibers used in our current experiment, such a system should be much less sensitive to environmental perturbations. 

With further advances in compact ion traps and optical lattices \cite{ivory2021integrated,  setzer2021fluorescence, niffenegger2020integrated, ropp2023integrating}, we anticipate that our dual-microcomb system, paired with an integrated atomic reference and compact narrow linewidth laser, may one day enable a fully-integrated high-performance optical atomic clock.

\vspace{\baselineskip}
\noindent\textbf{Methods}

\begin{figure*}[h]%
\centering
\includegraphics[width=1\textwidth]{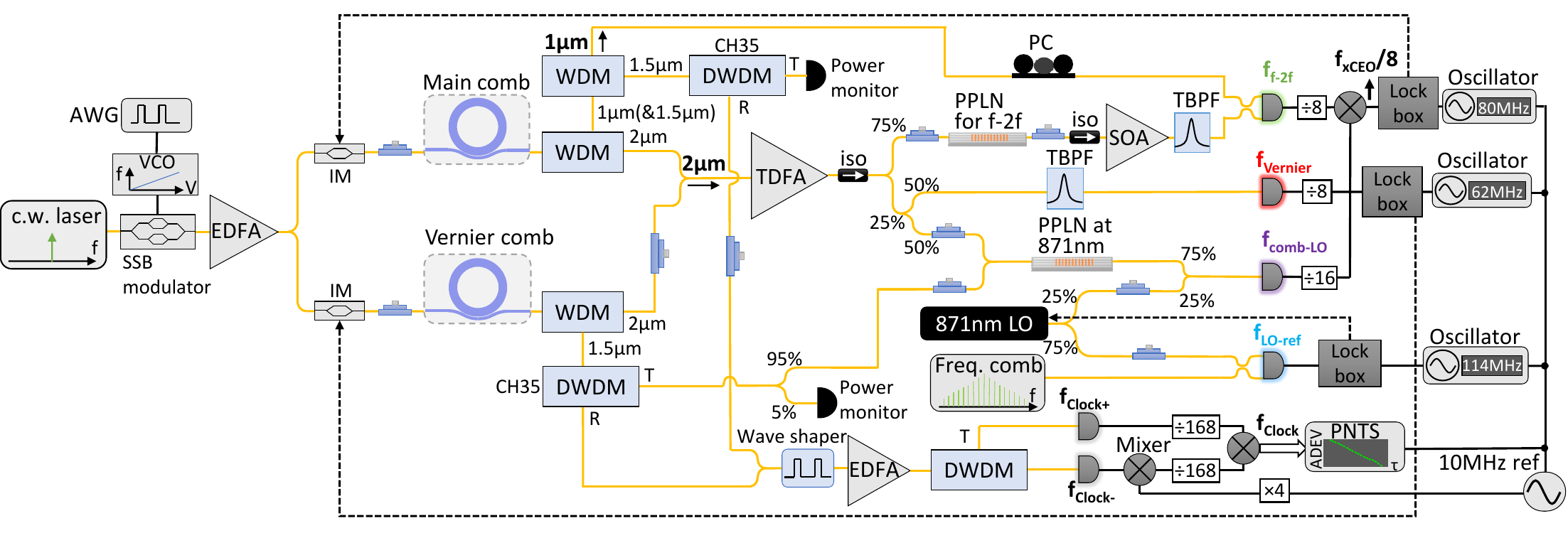}
\caption{Experimental setup. c.w.: continuous wave. AWG: arbitrary waveform generator. VCO: voltage-controlled oscillator. SSB modulator: single sideband modulator. EDFA: erbium-doped fiber amplifier. IM: intensity modulator. TDFA: thulium-doped fiber amplifier. PC: polarization controller. WDM: wavelength-division multiplexing. DWDM: dense wavelength-division multiplexing. iso: isolator. SOA: semiconductor optical amplifier. BPF: bandpass filter. TBPF: tunable bandpass filter. PNTS: phase noise test set}\label{figS1}
\end{figure*}
\noindent \small \textbf{Experimental Setup.} The experimental setup can be found in the Fig. \ref{figS1}. We use a single-sideband modulator to rapidly frequency-sweep \cite{stone2018thermal} a $\sim$1550 nm external cavity diode laser (ECDL) (Toptica CTL 1550) pump into resonance for both microrings simultaneously, generating microcombs in each. The two resonators are frequency-aligned using thermo-electric temperature control units. After comb generation is initiated, the pump frequency is adjusted through piezoelectric control of the ECDL towards a longer wavelength to optimize the dispersive wave conditions of the microcombs, so that the short wavelength dispersive wave of the main comb falls within the conversion wavelength of our available fiber-pigtailed PPLN waveguide (HC Photonics). In this dispersive wave state, we also make use of the long wavelength dispersive wave of the main comb, which benefits the SFG process for f-2f. A network of fiber optical filters and couplers are employed to separate spectral bands at $\sim$1550 nm, $\sim$2000 nm, and $\sim$1000 nm and combine the dual-comb spectra at $\sim$1550 nm and $\sim$2000 nm. For the results shown in the main text, we have introduced
foam partitions to cover most of our setup to reduce noise effects due to air currents and the like (further discussion in the Supplement).

The combined 2000nm spectral components are amplified using a thulium-doped fiber amplifier and then distributed in three arms for (a) the SFG process for f-2f, (b) the SFG process to hit 871nm via combining with the 1550nm residual pump from the Vernier comb, and (c) the Vernier beat detection. The SFG products for f-2f are subsequently amplified by a semiconductor optical amplifier (SOA, Innolume) and filtered by a 1nm-bandwidth bandpass filter (Photonwares) to suppress the broadband amplified spontaneous emission noise. It is then combined with the $\sim$1000 nm light from the main comb to generate the f-2f beat $f_{\rm f-2f}$ using a balanced photodetector (BPD). The SFG products at 871nm from the Vernier comb are combined with 25\% of the 871nm LO laser using a 75:25 coupler to generate $f_{\rm comb-LO}$, with $\sim$-36 dBm of SFG (spectrum shown in Fig. \ref{fig2}) and $\sim$-4.5 dBm of laser power measured at the photodetector. 75\% of the LO laser is combined with the fiber comb (bandpass-filtered at $\sim$871nm) using a 50:50 coupler and $f_{\rm LO-FC}$ is detected on another BPD, with $\sim$3 dBm of LO power and $\sim$-56 dBm power per fiber comb line at one arm of the BPD (where the laser and fiber comb spectra in Fig. \ref{fig2}(d) are measured). The clocks at both sides of the pump are selected by a programmable waveshaper (Finisar), optically amplified in the same EDFA, and separated with DWDM filters for detection on two distinct PDs.

The Vernier beat detected by an amplified PD is divided by 8 and sent to an offset phase lock servo (Vescent D2-135). The locked offset frequency is $N$ times the reference frequency generated by a stable oscillator (Agilent E8257D) synchronized to the 10 MHz GPS-disciplined oscillator (here, $N$=16). We obtain $f_{\rm xCEO}/8$ by mixing $f_{\rm f-2f}$ divided by 8 and $f_{\rm comb-LO}$ divided by 16. This beat is sent to another phase lock servo with $N$ set to 32. The reference signal is generated by another synchronized oscillator (Keysight 33220A).  
Finally, for our frequency instability measurements, we use frequency counters (Keysight 53230A) and a phase noise test set (Microsemi 5125A), which will be discussed in the next section.

\vspace{\baselineskip}
\noindent \textbf{Noise-Suppressed RF Clock.}
Our noise suppression scheme relies on mixing $f_{\rm clock+}$ and $f_{\rm clock-}$ to cancel differential-mode noise in the RF clock output. Due to the availability of components, we mix frequency-divided versions of these two signals in practice ($f_{\rm clock+}/168$ and $f_{\rm clock-}/168$). Importantly, the sum of these two signals is almost exactly equal to the second harmonic of either (that is, $f_{\rm clock+}/168 + f_{\rm clock-}/168 \approx 2f_{\rm clock+}/168 \approx 2f_{\rm clock-}/168$). This means it is impossible to filter out the undesirable second harmonics originating in the mixer and get the pure sum-frequency signal, thus still resulting in excess noise. To avoid this problem, we frequency shift one of the clocks by mixing with the 40 MHz signal (generated by twice-doubling the 10 MHz GPS-disciplined reference) to obtain $f_{\rm clock-}+40 \rm MHz$, filtered by a 30 MHz-bandwidth Yttrium Iron Garnet (YIG) bandpass filter. This frequency-up-shifted clock is being divided and electronically mixed with the other divided clock to realize $f_{\rm clock+}/168 + (f_{\rm clock-}+40 \rm MHz)/168$ at the output. The desired sum-frequency output is then distinguishable from harmonics of either of the mixer inputs. Although the second harmonic and other products still exist and are spaced by a $\sim$238 kHz (40MHz/168), the desired noise-suppressed output is $\sim$20 dB higher than the other products and the PNTS can track the desired signal (see Supplementary for RF spectrum).  Importantly, we note that the 10 MHz reference used here to shift one of the clock products should in principle be able to be derived from an optical reference if our system were to perform OFD on an atomic specimen (somewhat similar to \cite{diddams2001optical}). This would remove the need for a GPS-referenced 10 MHz sync signal in order to operate the OFD system.

For Fig. \ref{fig4}d we measure the frequency traces of the optical reference and the RF clock output simultaneously on two synchronized frequency counters running at 100 ms gate time with the same trigger signal and external gate. The optical reference frequency $f_{871}$ is obtained by recording $f_{\rm LO-FC}$ and calculating $f_{871}$ using Eq. \ref{eq LO-Menlo}. We attribute the larger frequency fluctuations in the RF clock trace compared with the optical reference to the spurious frequency content in the output signal. Although the weak spurs cause no significant problem for the PNTS (see Fig. \ref{fig4}a), they do result in a few times increase in the frequency fluctuations reported by the frequency counter. See the Supplement for further discussion.

The Allan deviation for the RF clock output and the fiber comb reference were taken several times over multiple days. The symbols and error bars plotted in Fig. \ref{fig4}a represent the averages and standard deviations from multiple measurement results. The error bars are relatively tight, indicating good repeatability.

\begin{figure}[h]%
\centering
\includegraphics[width=0.5\textwidth]{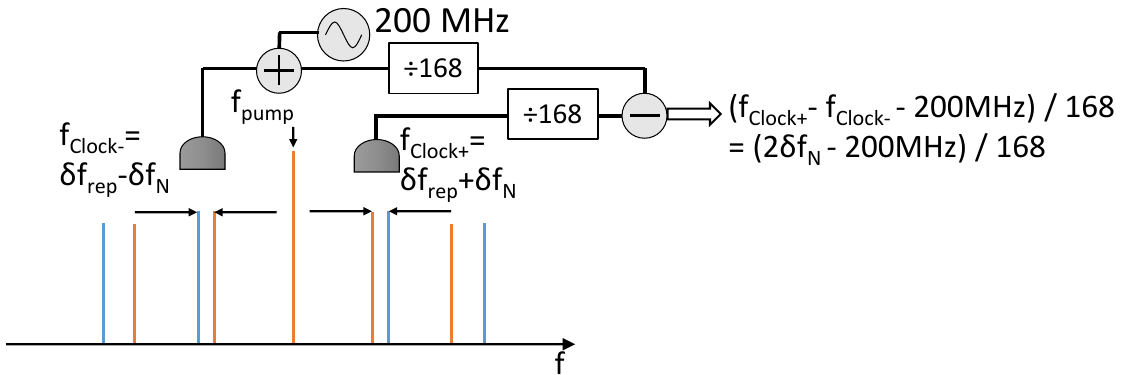}
\caption{Experimental configuration for interferometric noise extraction.}\label{figS4}
\end{figure}

\vspace{\baselineskip}
\noindent \textbf{Interferometric noise extraction.}
The noise suppression scheme can be modified to extract the interferometric noise for direct measurement. As explained in the main text, by adding $f_{\rm clock+}$ and $f_{\rm clock-}$ the differential frequency noise can be made to cancel out. By measuring the difference between $f_{\rm clock+}$ and $f_{\rm clock-}$, the clock term is cancelled and we obtain the frequency noise term $\delta f_{\rm N}(t)$. To implement the noise measurement, we frequency-up-shift one of the clocks by 200 MHz and divide it by 168, then frequency-mix with the other clock also divided by 168 (Fig. \ref{figS4}). This creates a frequency difference of $\sim$1.2 MHz out of the mixer, which is within the frequency range of our PNTS (1 - 400 MHz). The fractional instability of this signal filtered using a low-pass filter (normalized to the $\sim$235 MHz noise-suppressed clock frequency) is shown as the yellow curve in  Fig. \ref{fig4}a and is similar to the curve for the clock output without noise suppression.

\vspace{\baselineskip}
\noindent \textbf{Data Availability}

\noindent Data are available from the authors upon reasonable request.

\renewcommand{\bibfont}{\small}
\setlength{\bibsep}{0.0pt}
\bibliography{sn-bibliography}

\vspace{\baselineskip}
\noindent \textbf{Funding}

\noindent This work was funded in part by the DARPA APhI program, by AFOSR under grant FA9550-20-1-0283, and by the Swedish Research Council (VR 2020-00453). M.S.A. acknowledges support from the Researchers Supporting Project number (RSPD2023R613), King Saud University, Riyadh, Saudi Arabia.

\vspace{\baselineskip}
\noindent \textbf{Acknowledgements}

\noindent The effort at Purdue under the DARPA APhI program was part of a project team lead by Sandia National Laboratories. The Si$_3$N$_4$ fabrication was done at Myfab Chalmers. The authors thank Scott Diddams, Scott Papp, Frank Quinlan, Xu Yi, and Beichen Wang for fruitful discussion. Jizhao Zang suggested the experiment shown in Supplementary Figure S2 for diagnosis of the phase noise issue. Mike Kickbush and Mohammed Abu Khater assisted in identifying appropriate YIG filters for our system. Hayden McGuinness kindly reviewed the manuscript prior to submission. Some of our earlier work on Vernier dual combs, such as dual-octave comb pairs, was presented at CLEO 2021 (SW2H.7), CLEO 2022 (SW4O.2), and CLEO 2023 (STh1J.4 and STh1J.5).

\vspace{\baselineskip}
\noindent \textbf{Author Contributions}

\noindent A.M.W., M.S.A., and M.Q. devised the Vernier dual-comb scheme and initiated the project. K.W., N.P.O., and S.F. devised and conducted the experiments. C.W. designed the microring devices. M.G. and Z.Y. were responsible for fabrication of the microring devices, with oversight from V.T.-C..  D.E.L. was responsible for putting in place laboratory systems and advised on some of the experiments.  A.M.W. supervised the project. A.M.W., K.W., N.P.O., and S.F. wrote the manuscript.

\vspace{\baselineskip}
\noindent \textbf{Conflicts of Interest}

\noindent The authors declare no conflicts of interest.

\vspace{\baselineskip}
\noindent \textbf{Additional information}

\noindent More experimental details and investigations can be found in the Supplemental document. 

\end{document}


\title[Article Title]{Vernier Microcombs for Integrated Optical Atomic Clocks}


\author *[1]{\fnm{Kaiyi} \sur{Wu}}\email{\small wu1871@purdue.edu} \equalcont{\small These authors contributed equally to this work.}

\author[1]{\fnm{Nathan} \sur{P. O'Malley}} \equalcont{\small These authors contributed equally to this work.}

\author[1]{\fnm{Saleha} \sur{Fatema}}
\author[1,3]{\fnm{Cong} \sur{Wang}}
\author[2]{\fnm{Marcello} \sur{Girardi}}
\author[4]{\fnm{Mohammed} \sur{S. Alshaykh}}
\author[2]{\fnm{Zhichao} \sur{Ye}}
\author[1,5]{\fnm{Daniel} \sur{E. Leaird}}
\author[1]{\fnm{Minghao} \sur{Qi}}
\author[2]{\fnm{Victor} \sur{Torres-Company}}
\author[1]{\fnm{Andrew} \sur{M. Weiner}}

\affil[1]{\small\orgdiv{School of Electrical and Computer Engineering}, \orgname{Purdue University}, \orgaddress{\city{West Lafayette}, \postcode{47907}, \state{IN}, \country{USA}}}

\affil[2]{\small\orgdiv{Department of Microtechnology and Nanoscience}, \orgname{Chalmers University of Technology}, \orgaddress{\street{SE-41296},  \country{Sweden}}}

\affil[3]{\small\orgdiv{Currently with Department of Surgery}, \orgname{University of Pittsburgh}, \orgaddress{\postcode{15219}, \state{PA}, \country{USA}}}

\affil[4]{\small\orgdiv{Electrical Engineering Department}, \orgname{King Saud University}, \orgaddress{\city{Riyadh}, \postcode{11421}, \country{Saudi Arabia}}}

\affil[5]{\small\orgdiv{Currently with Torch Technologies}, \orgname{supporting AFRL/RW}, \orgaddress{\city{Eglin Air Force Base}, \state{FL}, \country{USA}}}

\maketitle


\section{Deriving Frequency Division Effect with Only Two Microcomb Locks}
As a result of the phase locks on $f_{\rm xCEO}$ and $f_{\rm Vernier}$, the repetition rates of the two microcombs are stabilized at the level of the optical reference. This results in a frequency division effect, as explained in the following. The clock frequency can be expressed in terms of the two microcomb beats being phase-locked ($f_{\rm xCEO}$ and $f_{\rm Vernier}$) to the LO laser using Eqs. 1, 2, \& 5: 
\begin{equation}
    f_{\rm clock} = \frac{1}{17292}(f_{\rm 871}-2f_{\rm xCEO}+385f_{\rm Vernier}).
     \label{eq freqdiv2}
\end{equation}

$f_{\rm 871}\sim344$ THz is an optical frequency, while the other two beats for locking are at the GHz level ($f_{\rm Vernier}\sim-8$ GHz and $f_{\rm xCEO}\sim$ 20.4 GHz). $f_{\rm 871}$ is roughly 4 and 2 orders of magnitude higher than $2f_{\rm xCEO}$ and $385f_{\rm Vernier}$, respectively. Importantly, in this work, the fiber comb optical reference and all RF references for the phase locks are stabilized to the same OCXO to eliminate relative drifting. Presuming good phase lock performance, such that the stability of the OCXO is transferred to the locked optical signals, the three terms on the right side of Eq. 11 will have similar fractional frequency instabilities. Given the multiple orders-of-magnitude differences in frequency of the terms in Eq. 11 and assuming good phase locks (hence similar fractional frequency instability), the clock output stability should be dominated by the stability of the optical frequency of the fiber comb-referenced LO laser. That is, $\Delta f_{\rm clock}(t) \approx \Delta f_{\rm 871}(t)/17292$, where $\Delta f(t)$ denotes frequency fluctuations in time of $f$, in Hz. In reality the phase locks may not function ideally but will simply need to be good enough not to corrupt the frequency instability too badly (by two orders of magnitude, in the case of $f_{\rm Vernier}$, and less stringent for the other lock). Note that if the Vernier beat frequency can be tuned closer to zero, the contribution of the Vernier term to the frequency instability of the clock will be reduced accordingly. In principle, the synchronizing signal for the phase locks here derived from the OCXO could instead be derived from the RF clock output. This would allow referencing of these locks to an atomic reference in an actual optical atomic clock system.

We can further consider the third phase lock ($f_{\rm LO-FC}$) using Eqs. 6 and 11:
\begin{equation}
    f_{\rm clock} = \frac{1}{17292}(f_{\rm FC}+f_{\rm LO-FC}-2f_{\rm xCEO}+385f_{\rm Vernier}),
\end{equation}
which reveals the requirements for locking the 871 nm laser to the fiber comb by a similar line of reasoning. Since $f_{\rm LO-FC}$ is roughly five orders of magnitude lower than the 344 THz LO, this is the least demanding of the three locks. 

In summary, we expect frequency division from optical-to-RF with only a small contribution from the phase locks, even despite the free-running $f_{\rm CEO}$.

\section{Estimating Optical Stability of Fiber Comb Modes}
\begin{figure}[h]%
\centering
\includegraphics[width=0.9\textwidth]{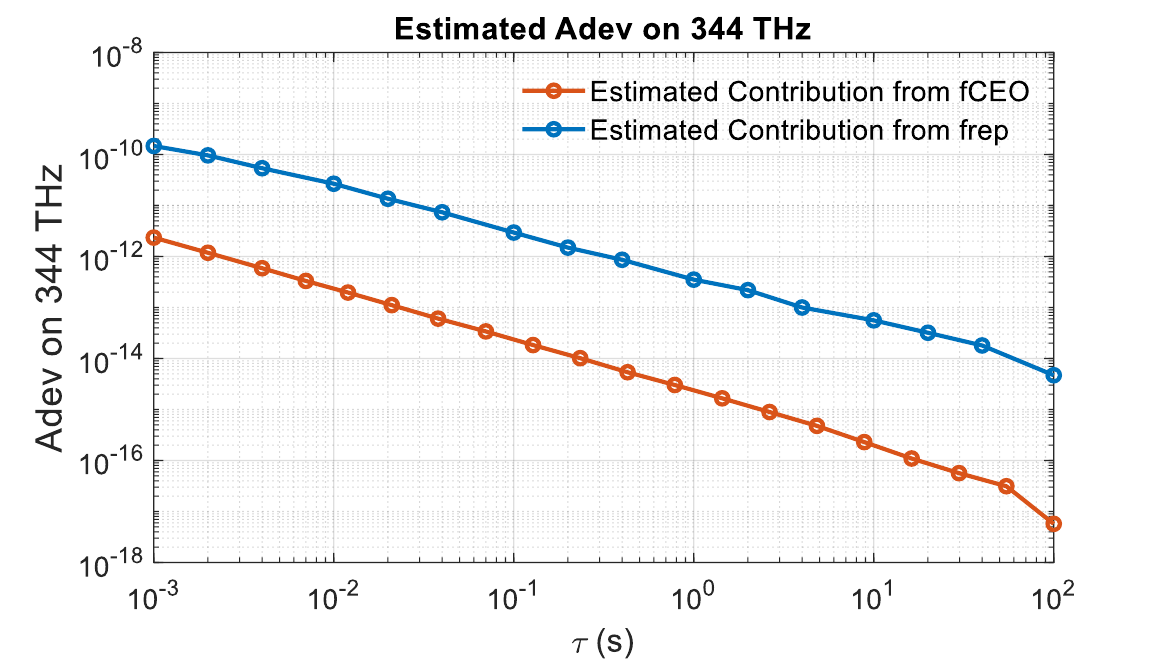}
\caption{Estimated relative contributions of the repetition rate and CEO frequency to the 344 THz optical reference line.}\label{figS5}
\end{figure}

As we are using one of the fiber comb lines as the optical reference, in this section we estimate its frequency instability. The frequency instability of any given fiber comb mode can be straightforwardly seen through the typical frequency comb equation ($f_m = m\times f_{\rm rep} + f_{\rm CEO}$) to depend on the two terms $m\times f_{\rm rep}$ and $f_{\rm CEO}$. In lieu of measuring the actual optical stability of a comb mode directly (which would require a second stable optical reference not available in our laboratory), one can measure the stability of each of the two terms separately to form an optical stability estimate. The fractional frequency instability of the $m\times f_{\rm rep}$ term can be easily obtained by measuring the repetition rate at 250 MHz, as described in the main text. The fractional instability of $m\times f_{\rm rep}$ should then ideally be the same as $f_{\rm rep}$. The stability of the $f_{\rm CEO}$ term is estimated by sending an in-loop monitor output of the fiber comb offset frequency to a frequency counter referenced to the GPS-disciplined oscillator. The resulting frequency instabilities from the two measurements can be seen in figure \ref{figS5}, clearly indicating that the repetition rate contribution is expected to dominate the CEO frequency contribution. Thus we estimate the optical mode's fractional frequency instability simply as that of the repetition rate.

\section{Interferometric noise investigation}

\begin{figure}[h]%
\centering
\includegraphics[width=0.9\textwidth]{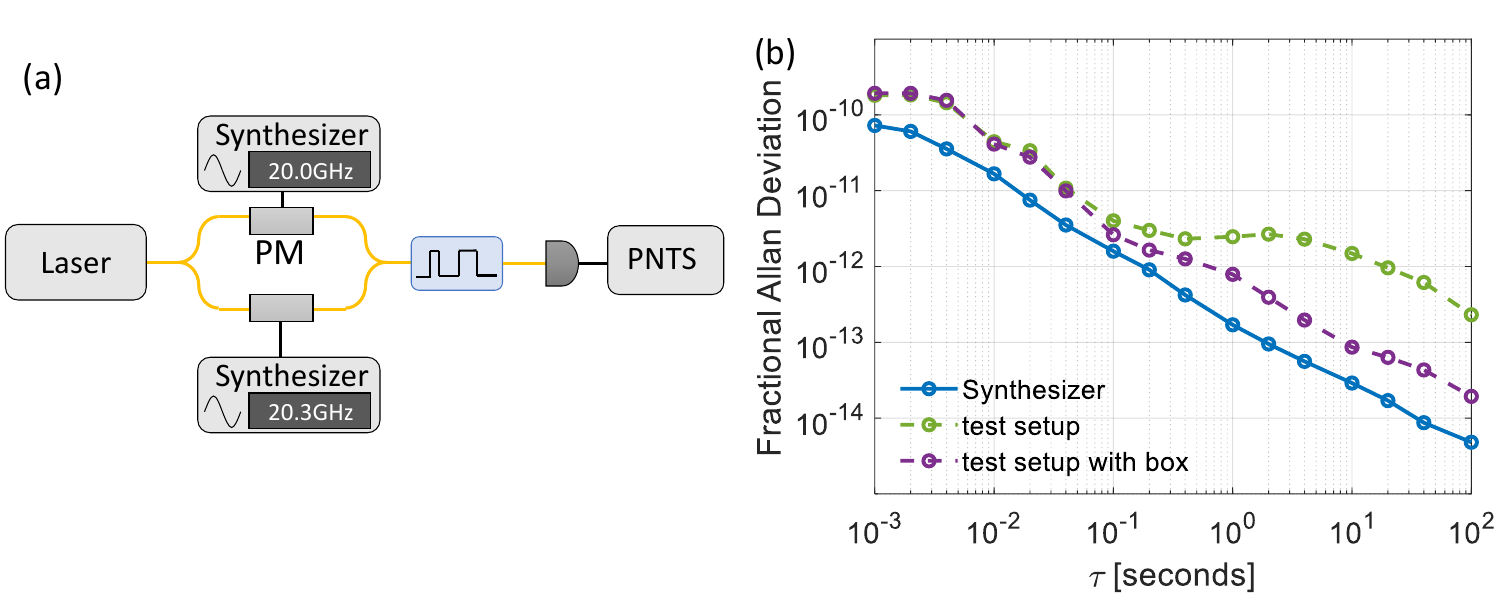}
\caption{Investigation on the interferometric noise using an analogous test setup. \textbf{a}, the schematic of the test setup. \textbf{b}, Fractional Allan deviations of the heterodyne beat between the first sidebands generated by two phase modulators with (purple line) and without (green line) a cover, and the frequency reference (blue line).}\label{figS2}
\end{figure}

To gain insight into the interferometric noise in our system, we investigate a simple fiber interferometer setup with a stable heterodyne beat frequency as shown in Fig. \ref{figS2}a, analogous to our dual-comb configuration as a proof-of-concept experiment. Here we place a phase modulator (analogous to the microring) in each arm of a fiber Mach-Zehnder interferometer. The two phase modulators are driven by two synthesizers at 20 GHz and 20.3 GHz, respectively.
We optically filter the first sideband from both arms and measure the frequency stability of the heterodyne beat note  (analogous to the clock beat in the microcomb system). The two synthesizers are synchronized to a common 10 MHz clock, so in absence of optically-introduced noise, we expect this signal to be quite frequency-stable, and average down as $\sim 1 / \tau$. However, we again observe a plateau feature in the Allan deviation in this measurement, closely resembling the plateau feature in the clock in the dual-microcomb system discussed in the main text (see Fig. \ref{figS2}b). This experiment indicates that the plateau feature of our clock signal is due to fiber-related noise in the interferometer-like segment of our microcomb system. 
To mitigate the environmental effects, we cover the interferometer setup in Fig. \ref{figS2}a with a foam box, which can block most of the airflow in the laboratory. The Allan deviation improves significantly and no longer shows an obvious plateau feature, although there is still excess frequency noise (Fig. \ref{figS2}b). 

\begin{figure}[h]%
\centering
\includegraphics[width=0.9\textwidth]{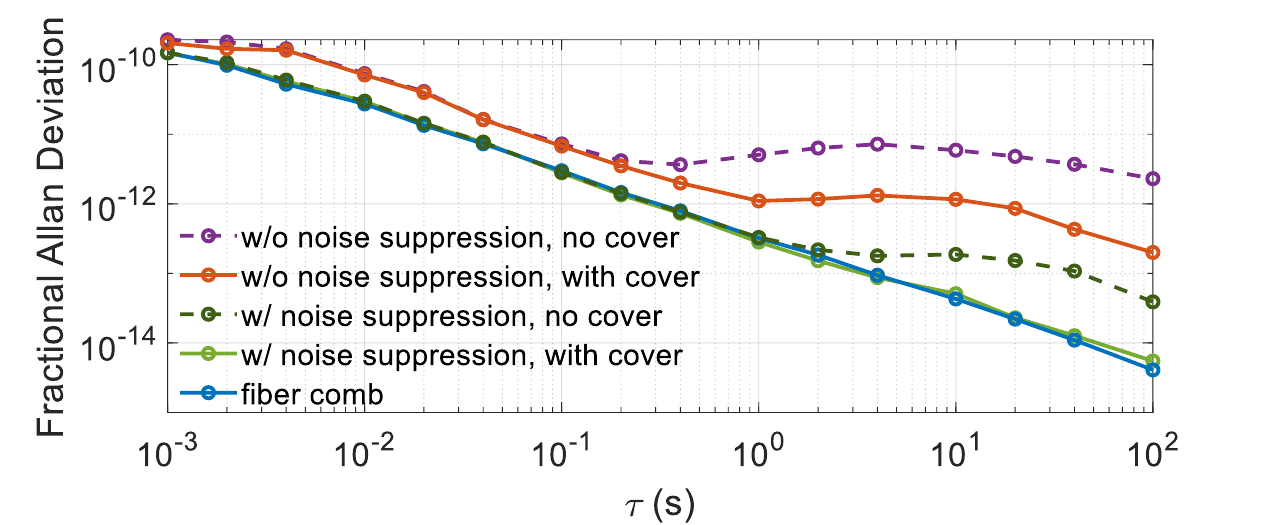}
\caption{Fractional Allan deviations of the clock output without and with the effects of noise-suppression scheme and the covers, and the fiber comb optical reference.}\label{figS3}
\end{figure}

We also further investigate the interferometric noise effects in the dual-comb setup and attempt to estimate the noise suppression ratio achieved.  Figure \ref{figS3} shows the measured clock instabilities with and without foam partitions covering the setup and with and without noise suppression.  The orange and green solid curves are the instabilities with the foam covers in place, without and with our noise suppression scheme, respectively, as reported in the main text.  Without noise suppression, there is excess frequency instability at short averaging times and a plateau starting around 1 s average time.  With noise suppression, the fractional frequency instability is brought down to that of the optical reference, which constitutes our measurement floor.  The difference between the orange and green solid curves suggests $\sim$30-40$\times$ improvement; however, since we reach our measurement floor, this estimate is a lower bound.  To gain additional information, we remove part of the foam cover and again compare the fractional frequency instability without and with noise suppression (purple and green dashed lines, respectively).  This raises the noise further above that of the optical reference and provides an increased dynamic range for measurement. Here we estimate 40× noise suppression in the plateau region.  A simple theoretical estimate of our noise suppression scheme (see the last section of Supplementary) based on the slight wavelength difference between the participating comb lines for $f_{\rm clock+} \& f_{\rm clock-}$ suggests that ideally, the noise suppression capability should be of the same order of magnitude but somewhat higher.  The difference may arise from noise or nonlinearities in our electronic mixing and division scheme.

\section{Frequency Counter Resolution Limitations}

\begin{figure}[h]%
\centering
\includegraphics[width=0.75\textwidth]{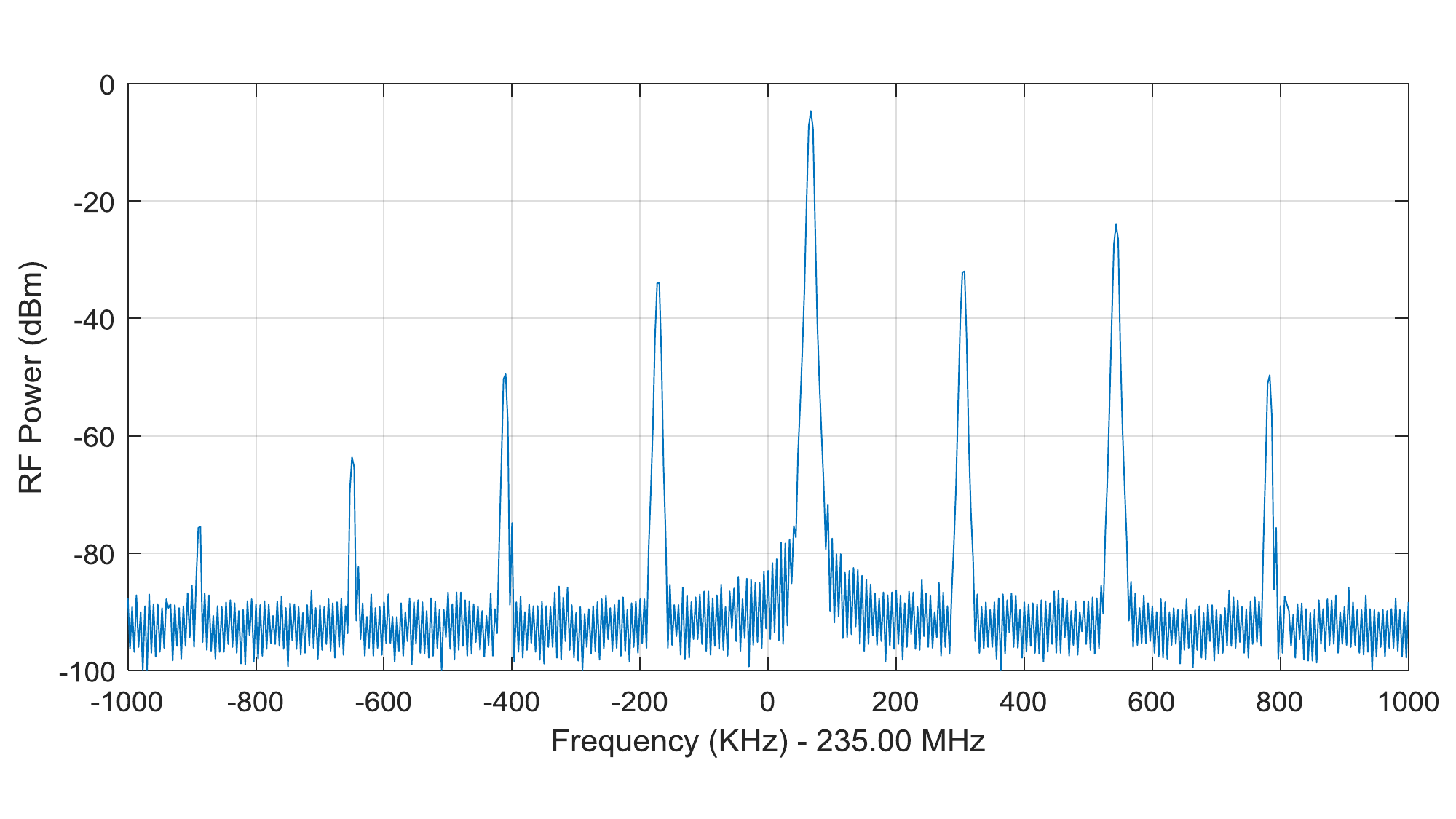}
\caption{Electrical spectrum analyzer trace of noise-suppressed clock output at 3 kHz RBW.}\label{figS7}
\end{figure}

\begin{figure}[h]%
\centering
\includegraphics[width=1\textwidth]{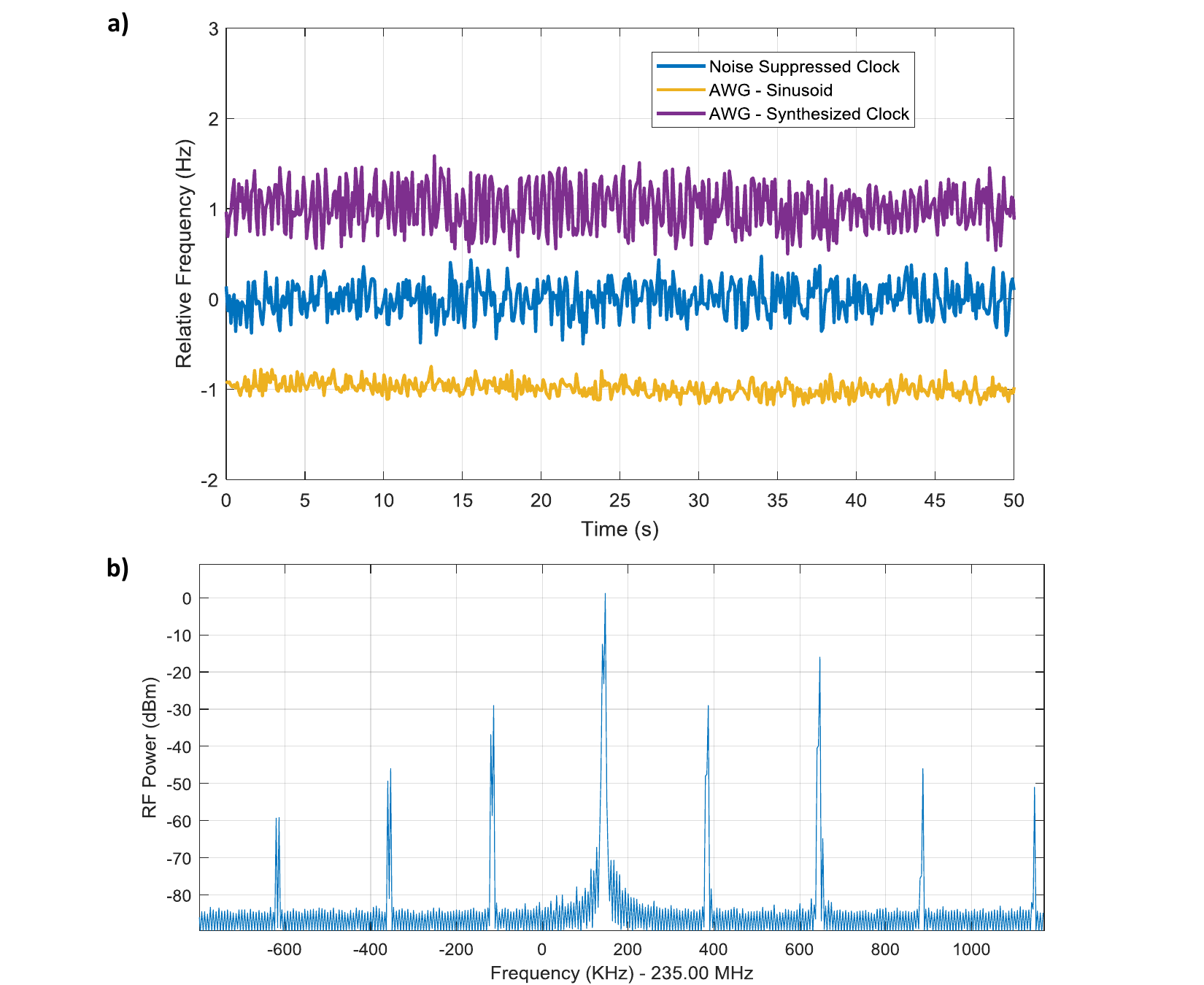}
\caption{\textbf{a}, counter measurements of the $\sim$235 MHz noise suppressed clock signal (blue trace), $\sim$235 MHz sinusoid generated by AWG (yellow trace), and the AWG-synthesized clock signal (purple trace). \textbf{b}, Electrical spectrum analyzer trace of AWG-synthesized clock signal at 1 kHz RBW.}\label{figS8}
\end{figure}

As noted in the main text, Fig. 4e (the time trace of the noise-suppressed clock output) shows a wider trace width than is seen in Fig. 4d (the time trace of the 871 nm laser). We attribute this to resolution limitations of our frequency counters. In order to support our claim, we observe that, as mentioned in the main text, our noise-suppressed clock output is not a pure sinusoid. Due to the nonlinearity of the mixer combining $f_{\rm clock+}$ and $f_{\rm clock-}$, there are spurious tones in the output spectrum, as can be seen in Fig. \ref{figS7}. The spurs are separated by $\sim$240 kHz, as expected, due to the 40 MHz shift applied to $f_{\rm clock-}$ and the electrical division factor of 168. In addition to the typical counter timing jitter-related resolution limits, these spurious tones can cause extra noise in the frequency counter measurement. 

We confirm this by testing one of our frequency counters with synthesized waveforms from an arbitrary waveform generator (AWG). Initially, a pure sinusoidal tone is generated at $\sim$235 MHz, the frequency of the noise-suppressed clock, and sent to the frequency counter at 100 ms gate time (driven by an external gate signal). The data are plotted in Fig. \ref{figS8}a (yellow trace). The resulting trace appears cleaner than the clock output (plotted here in the blue trace for comparison) and is approximately at the resolution limit of the frequency counter. Note the traces have been shifted in post-processing for easier comparison. The AWG is then programmed to generate a waveform with a spectrum nearly identical to that of the clock (see Fig. \ref{figS8}b), and the counter trace becomes noisier as a result (purple trace). This effect is observed with a flat phase on each tone in the spectrum, as well as in two further trials with randomized phases applied to each line when programming the AWG (not shown here). This indicates the additional spectral content increases noise in the frequency counter measurement.

In summary, in addition to noise introduced by the clock being near the timing jitter-induced resolution limit of the frequency counter, the spurious signals in the noise-suppressed clock (caused by nonlinearities in our electronic mixer) induce further degradation in the counter measurement of the clock. We attribute the differences between Figs. 4d and e to these causes.

\section{Theoretical analysis of noise suppression}

\begin{figure}[h]%
\centering
\includegraphics[width=0.5\textwidth]{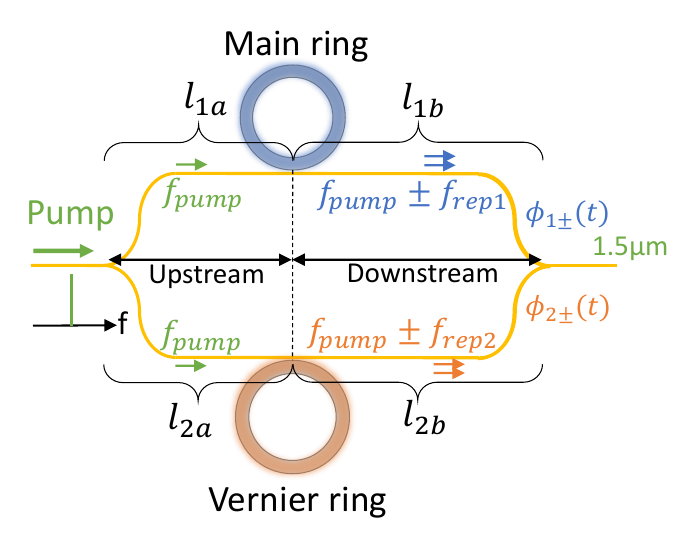}
\caption{Schematic illustration of the dual-comb setup notated with physical parameters.}\label{figS6}
\end{figure}

We theoretically analyze the effectiveness of our noise suppression scheme with the participating comb lines for $f_{\rm clock+} \& f_{\rm clock-}$ being at slightly different frequencies. Fig. \ref{figS6} shows a simplified dual-comb schematic notated with the physical parameters involved in the analysis. $l_{\rm 1a}$ and $l_{\rm 2a}$ denote the fiber lengths between the splitting of the pump and the main and Vernier microrings, respectively. $l_{\rm 1b}$ and $l_{\rm 2b}$ denote the fiber lengths from the main and Vernier rings to where the fibers are combined. We assume the phase fluctuations in the fibers result from fiber length variations, and the refractive index $n$ for the participating comb lines in the fibers remain constant for simplification in the analysis. The fiber length with phase fluctuations can thus be written as follows:
  \begin{equation}
  l_{\rm x}(t) = \Bar{l}_{\rm x} + \delta l_{\rm x}(t)
  \label{eq length}
  \end{equation}
which consists of the average value $\Bar{l}_{\rm x}$ that is constant in time, and the time varying length $\delta l_{\rm x}(t)$ contributed by the fiber noise. The subscript x denotes 1a, 2a, 1b, 2b, the four segments of the fiber length of interest. 

The phase fluctuation $\phi(t)$ for the participating comb lines of $f_{\rm clock+}$ and $f_{\rm clock-}$ can be expressed in terms of fiber length:
\begin{equation}
    \begin{split}
    & \phi_{1+}(t) = \frac{2 \pi f_{\rm pump} n}{c} l_{\rm 1a}(t) + \frac{2 \pi (f_{\rm pump} + f_{\rm rep1}) n}{c} l_{\rm 1b}(t), \\
    & \phi_{1-}(t) = \frac{2 \pi f_{\rm pump} n}{c} l_{\rm 1a}(t) + \frac{2 \pi (f_{\rm pump} - f_{\rm rep1}) n}{c} l_{\rm 1b}(t), \\
    & \phi_{2+}(t) = \frac{2 \pi f_{\rm pump} n}{c} l_{\rm 2a}(t) + \frac{2 \pi (f_{\rm pump} + f_{\rm rep2}) n}{c} l_{\rm 2b}(t), \\
    & \phi_{2-}(t) = \frac{2 \pi f_{\rm pump} n}{c} l_{\rm 2a}(t) + \frac{2 \pi (f_{\rm pump} - f_{\rm rep2}) n}{c} l_{\rm 2b}(t),
    \end{split}
    \label{eq phis}
\end{equation}
where we use $\phi_{1+}(t)$ and $\phi_{2+}(t)$ for the higher-frequency sidebands for the main and Vernier comb line, respectively, and similarly, we use $\phi_{1-}(t)$ and $\phi_{2-}(t)$ for the lower-frequency sidebands. 

For the noise-suppressed clock, we modify Eqs. 8 \& 9 in the main text by considering the wavelength dependence of $\phi_1$ and $\phi_2$ using the above equations (Eq. \ref{eq phis}) and obtain: 
\begin{equation}
    f_{\rm clock+} + f_{\rm clock-} = 2 (f_{\rm rep1} - f_{\rm rep2}) + \frac{1}{2\pi} (\phi'_{1+} - \phi'_{2+} + \phi'_{2-} - \phi'_{1-}). 
    \label{eq clockphi}
\end{equation}
Thus, the additional frequency instability under the noise suppression scheme is from the time-varying phase terms. Substituting the expressions of Eq. \ref{eq phis} into Eq. \ref{eq clockphi}, we obtain the residual frequency noise for the noise-suppressed clock:
\begin{equation}
\begin{split}
    \delta (f_{\rm clock+} + f_{\rm clock-}) & = \frac{n}{c} [2f_{\rm rep1}\delta l'_{\rm 1b}(t)-2f_{\rm rep2}\delta l'_{\rm 2b}(t)] \\
    &\approx \frac{n}{c}2f_{\rm rep}[\delta l'_{\rm 1b}(t) - \delta l'_{\rm 2b}(t)].    
\end{split}
\end{equation}
$l'_{\rm x}(t)$ is the time derivative of $l_{\rm x}(t)$. The time derivative of phase fluctuation $\phi'(t)$ is from the fiber length fluctuation $l'_{\rm x}(t)$. Using the relation in Eq. \ref{eq length} and since $\Bar{l}_{\rm x}$ is constant, we have $l'_{\rm x}(t) = \delta l'_{\rm x}(t)$. Additionally, since the two microcomb repetition rates differ by only $\sim$ 2\%, we approximate them as $f_{\rm rep1} \approx f_{\rm rep2} \approx f_{\rm rep}$. 

The fractional frequency instability for the noise-suppressed clock is:
\begin{equation}
    \frac{\delta (f_{\rm clock+} + f_{\rm clock-})}{2f_{\rm clock}} \approx \frac{n}{c} \frac{f_{\rm rep}}{f_{\rm clock}}[\delta l'_{\rm 1b}(t) - \delta l'_{\rm 2b}(t)]. 
    \label{eq fracmixedclock}
\end{equation}
This equation indicates that ideally the frequency noise remaining after the suppression scheme depends only on fiber length fluctuations downstream from the microrings; in the ideal case length fluctuations experienced by the pump upstream from the microrings cancel out completely. Note that care is needed in interpreting the subtraction operation on the right hand side of eq. \ref{eq fracmixedclock}.  If the length fluctuations are highly correlated and in phase, they will tend to cancel each other out.  However, if the length correlations are uncorrelated, their effects add incoherently. 


For comparison, we also deduce the fiber-related frequency noise without the noise suppression. The clock without noise suppression can be written as:
\begin{equation}
    f_{\rm clock+} = f_{\rm rep1} - f_{\rm rep2} + \frac{1}{2\pi} (\phi'_{1+} - \phi'_{2+}). 
    \label{eq oneclockphi}
\end{equation}
The fiber-related frequency noise will be: 
\begin{equation}
\begin{split}
    \delta f_{\rm clock+} &= \frac{n}{c} f_{\rm pump} (\delta l'_{\rm 1a} - \delta l'_{\rm 2a}) + \frac{n}{c} [(f_{\rm pump}+f_{\rm rep1})\delta l'_{\rm 1b}-(f_{\rm pump}+f_{\rm rep2})\delta l'_{\rm 2b}] \\
    & \approx \frac{n}{c} f_{\rm pump} (\delta l'_{\rm 1a} - \delta l'_{\rm 2a} + \delta l'_{\rm 1b} - \delta l'_{\rm 2b}).
    \label{eq oneclocknoise}
\end{split}
\end{equation}
Since the repetition rate is only 0.5\% of the pump frequency, in the second line of Eq. \ref{eq oneclocknoise} we consider the sideband frequencies approximately equal to the pump frequency. 
In this approximation $\delta f_{\rm clock-}\approx \delta f_{\rm clock+}$. We thus use $\delta f_{\rm clock}$ to represent either clock signal interchangeably. The fiber-related frequency noise for $f_{\rm clock}$ depends on the pump frequency and the phase fluctuation both upstream and downstream of the microrings. Using $l_{\rm j}=l_{\rm ja}+l_{\rm jb}$ and $\delta l'_{\rm j} = \delta l'_{\rm ja}+\delta l'_{\rm jb}$ (j = 1,2), we can express the fractional frequency instability as:

\begin{equation}
    \frac{\delta f_{\rm clock}}{f_{\rm clock}} \approx \frac{n}{c} \frac{f_{\rm pump}}{f_{\rm clock}} (\delta l'_{\rm 1} - \delta l'_{\rm 2}). 
    \label{eq fraconeclock}
\end{equation}

To characterize the effectiveness of the noise-suppression scheme, we define a parameter $\eta$ equal to the ratio of the fractional frequency instability without (Eq. \ref{eq fraconeclock}) and with (Eq. \ref{eq fracmixedclock})  noise-suppression.

\begin{equation}
    \eta = \frac{f_{\rm pump}}{f_{\rm rep}} \frac{\delta l'_{\rm 1} - \delta l'_{\rm 2}}{\delta l'_{\rm 1b} - \delta l'_{\rm 2b}} = \eta_{f}\eta_{l}. 
    \label{eta equation}
\end{equation}
Thus, in theory, the noise-suppression ratio is the product of two factors. The first is the frequency-dependent term $\eta_{f}=f_{\rm pump}/f_{\rm rep}$, which is $\sim$200 in our experiment. The second factor $\eta_{l}$ depends on the time-dependent fiber length fluctuations, including the correlations (or lack of correlations) between them. The behavior of $\eta_{l}$ depends on the specific experimental layout and environment perturbations.

\newpage